**A Market for Lemons? Strategic Directions for a Vigilant Application of Artificial Intelligence in Entrepreneurship Research**

Martin Obschonka§
Amsterdam Business School
University of Amsterdam
Plantage Muidergracht 12, 1018 TV Amsterdam
1001 NL Amsterdam
m.obschonka@uva.nl

Moren Lévesque
Schulich School of Business
York University
4700 Keele Street
Toronto, Ontario, Canada M3J 1P3
mlevesque@schulich.yorku.ca

§ Corresponding author

*Acknowledgements*: The authors acknowledge highly constructive comments from Evan Douglas, Christian Fisch, Michael Fritsch, Pierre Gautreau, Samuel D. Gosling, and Jackson Lu on earlier versions of this manuscript, as well as from scholars participating in the 2024 Rencontres de St-Gall, St. Gallen, Switzerland.

**ABSTRACT**

The rapid expansion of AI adoption (e.g., using machine learning, deep learning, or large language models as research methods) and the increasing availability of big data have the potential to bring about the most significant transformation in entrepreneurship scholarship the field has ever witnessed. This article makes a pressing meta-contribution by highlighting a significant risk of unproductive knowledge exchanges in entrepreneurship research amid the AI revolution. It offers strategies to mitigate this risk and provides guidance for future AI-based studies to enhance their collective impact and relevance. Drawing on Akerlof's renowned market-for-lemons concept, we identify the potential for significant knowledge asymmetries emerging from the field's evolution into its current landscape (e.g., complexities around construct validity, theory building, and research relevance). Such asymmetries are particularly deeply ingrained due to what we term the double-black-box puzzle, where the widely recognized black box nature of AI methods intersects with the black box nature of the entrepreneurship phenomenon driven by inherent uncertainty. As a result, these asymmetries could lead to an increase in suboptimal research products that go undetected, collectively creating a market for lemons that undermines the field's well-being, reputation, and impact. However, importantly, if these risks can be mitigated, the AI revolution could herald a new golden era for entrepreneurship research. We discuss the necessary actions to elevate the field to a higher level of AI resilience while steadfastly maintaining its foundational principles and core values.

**Keywords:** actionable guidelines, artificial intelligence (AI), asymmetric information, entrepreneurship scholarship, knowledge exchanges, research processes

> "The greatest danger in times of turbulence is not the turbulence; it is to act with yesterday's logic."
> —**Peter F. Drucker** (1909 to 2005)
> Management guru, educator, and author

## 1. Introduction

The major impact of the artificial intelligence (AI) revolution on scientific research and knowledge development is indisputable and profound. Advanced AI methods[1] such as machine learning, deep learning, and large language models[2] are transforming entire research fields.[3] One can also distinguish, for instance, between predictive AI and generative AI (Hermann & Puntoni, 2024). Predictive AI is a powerful research method that (seemingly) excels in tasks such as pattern recognition within existing data to predict future outcomes (Blohm et al., 2022; Matz & Netzer, 2017). In contrast, generative AI has a (seemingly) impressive capacity for creating new content (Susarla et al., 2023). With this technological progress in the AI space has come new challenges, uncertainties, and questions for scholars regarding the use of AI and the need for respective guidelines and regulation (European Commission, 2024; Gatrell et al., 2024).

This article focuses on the potential near-term impact of the AI disruption on the field of entrepreneurship research and its stakeholders. The topic of AI, including AI-supported research (e.g., using predictive and/or generative AI as a research method), has recently become prominent in entrepreneurship scholarship. This is evidenced by various journal special issues (including special issue projects in *Entrepreneurship Theory and Practice* with Obschonka et al., 2024; in *Journal of Business Venturing* with Ramoglou et al., 2024; or in *Small Business Economics Journal* with Obschonka & Audretsch, 2020), editorials (e.g., Lévesque et al., 2022), and numerous commentaries and reflections (e.g., Chalmers et al., 2021; Lévesque & Stephan, 2020; Maula & Stam, 2020; Schwab & Zhang, 2019; Shepherd & Majchrzak, 2022; Wennberg & Anderson, 2020). Entrepreneurship scholars have begun to debate how AI will impact the field; for example, whether AI can replace human scholars in literature review tasks (Robledo et al., 2023) or how to best employ generative AI and prompting in the research process as an augmentation for human scholars (Ferrati et al., 2024). Therefore, this article offers a timely meta-contribution by analyzing the current state of the entrepreneurship field in light of the unfolding AI revolution. Importantly, we diagnose far-reaching risks of unproductive knowledge exchanges. By identifying these risks and proposing strategies to mitigate them, we provide a roadmap for the future wave of AI-based studies in the field, to enhance their impact and relevance. Our work serves as a field-wide reflection and warning, aiming to elevate the collective contribution of AI research in entrepreneurship.

[1] Our definition of advanced AI methods is the increasingly complex, difficult to understand (for human), and unexplainable applications of tools that are advancing, progressing, and getting adopted in the research processes.
[2] Large language models (LLMs) are artificial intelligence (AI) programs that can recognize and generate text, and can 'read' and process massive data sets.
[3] For social science scholars, a fundamental difference between advanced AI methods like machine learning algorithms and traditional empirical methods like regressions, is the loss of control over statistical assumptions and the conscious selection of independent variables based on theory. Traditional empirical methods help us avoid erroneous correlations and provide mechanisms to detect certain statistical errors. However, the dubious practice of p-hacking remains possible (and is sometimes even expected to increase) with advanced AI methods.

It is safe to say that the contemporary scholarly field of entrepreneurship, which is largely anchored in the initial explosion stage of the 2000s, coined the field's 'golden era' (Landström, 2020), is now on the brink of a new major transformation: the 'explosion' of a completely new generation of entrepreneurship research powered by the massive AI wave that has reached most scholarly fields.[4] Through its evolution, the entrepreneurship field is likely to be hit by this unfurling wave in a unique way relative to other fields, due to six major triggers—complexities around construct validity, theory building, research relevance, publication pressure, education/training, and inter-, multi-, and trans-disciplinarity—when using advanced AI methods. As explained in more detail later, these triggers are further complicated by what we term the double-black-box puzzle for the field, where the widely recognized black box nature of AI methods (Klauschen et al., 2024; Messeri & Crockett, 2024; Rai, 2020) intersects with the black box nature of the entrepreneurship phenomenon that is driven by its inherent uncertainty. In other words, human scholars could struggle to understand and validate a black-box research method applied to a black-box phenomenon.

Highlighting these dynamics, we see the potential for significant unfolding information asymmetries—the key concern in this article—inspired by Akerlof's (1970) focus on such asymmetries in his market-for-lemons concept. Articles using AI methods might appear high-quality (e.g., due to the apparent sophistication and innovativeness of the advanced AI techniques and big data), but if the authors (and readers like reviewers) cannot fully understand the methods' and results' limitations, they may accept suboptimal or even flawed research as optimal and valid, much like Akerlof's famous example of used cars in a market where flaws are undetectable to potential car buyers. Over time, this could lower the overall quality of research in the field (gravitating toward undetected suboptimal research). This view is also motivated by entrepreneurship scholarship, unlike other related fields (e.g., psychology, sociology, and economics), having a legacy of being sometimes behind the curve in terms of pursuing methodological research innovations, probably for understandable reasons. The field's chief capability has been the central subject—the fascinating phenomenon of entrepreneurship and its fundamental role in the economy and wider society. The field, rightfully so, claims and further strives to understand this complex, real-world phenomenon and its underlying drivers, which may direct the research focus of entrepreneurship scholars primarily toward this idiosyncratic phenomenon, potentially with little incentive to produce methodological innovations that could also be pioneers in other fields. However, we also acknowledged the great ambitions to advance methods in the field (e.g., recent method-oriented editorials in major entrepreneurship journals).

[4] Our article is not legal advice, and it should not be considered the journal or publisher's official position regarding the use of AI, including generative AI, in the research or review process.

We see major vulnerabilities to field-wide market-for-lemons dynamics (Akerlof, 1970), in which the lemons stand for difficult-to-detect suboptimality in research products using AI methods and disseminated in academic journals, and the market dynamics are the resulting unproductive knowledge exchanges (and their negative broader impact on entrepreneurship scholarship and its stakeholders). Therefore, the potential detrimental effects of unproductive knowledge exchanges for the field should not be underestimated as they could not only establish such a market, but also reinforce mechanisms that maintain and even amplify such a market (e.g., basing new scholarship such as new research and theories—and the education and training of the next generation of entrepreneurship scholars—on such findings). Hence, such market-for-lemons dynamics could obstruct the positive upward trajectory that the field has enjoyed over the past decades, particularly in terms of becoming a *bona fide* field with a strong reputation and scientific rigor—a field that can provide "solutions, or at least the promise of solutions to some of society's most compelling problems" (Audretsch, 2021:852). Our central point is that this positive trajectory deserves to be protected, also in the new AI era. Hence, our emphatic warning in this article.

We recognize that other research fields also face unique triggers flowing from AI's entry into research and publishing, but the entrepreneurship field can face *particular* triggers that, also via their unique interplay, can result in a growing market for lemons in this field (and it may be difficult to be fully aware when such a market actually emerges and persists, as it can also go undetected). We thus argue that we must, as an academic community, accelerate structural changes in the entrepreneurship research field by proactively shaping the research infrastructures and processes, and setting effective regulations and standards.[5] These should ensure that the field does not become a passive recipient of AI adoption but instead moves quickly ahead of the anticipated massive adoption curve by proactively creating AI resilience *before* massive adoption takes place. In other words, under the AI revolution—using Peter Drucker's expression in the introductory quote—'yesterday's logic' may no longer apply. We might need a new type of thinking in our research field. Part of this should be proactive preparation for the transformation, which has the capacity to essentially weed out the lemons and, instead, create something very different like a market for 'diamonds', if one wishes to put it that way.[6] Diamonds would be verifiable-high-quality research products using AI methods disseminated in academic journals. Producing a wealth of such diamonds, and avoiding lemons that circulate undetected, could help stimulate unprecedented growth and prosperity in the field by ensuring that AI methods are used to produce and disseminate optimal, rather than potentially suboptimal, research products. As a result of the AI revolution, the entrepreneurship field would then even increase its relevance for society, and therefore also invigorate entrepreneurship research fueled by the AI revolution. In this positive, somewhat enthusiastic scenario, it could result in a second 'golden era' for the field (Landström, 2020).

Therefore, we offer some guidelines to help guide our field, raising awareness of the potential for a market for lemons and suggesting concrete action points for various stakeholders, both individually and collectively. This is a pressing issue, given that the unfolding AI transformation of the field is probably one of the most disruptive forces it has ever witnessed (Obschonka & Audretsch, 2020), shaking its very foundations. In this vein, we concur with Grimes et al. (2023), who indicate that deep AI disruptions might compel the entrepreneurship field to reflect on and renew its core identity and justification for existence. This might be a somewhat challenging and arduous process, but collectively (i.e., involving key stakeholders such as authors, reviewers, editors, and journal publishers) it is achievable. We emphasize that we are unwavering in our general support for the adoption of advanced AI methods within ethical frameworks (e.g., Lévesque et al., 2022; Obschonka & Fisch, 2022), while simultaneously issuing a stern warning about the potential hidden challenges they may present for achieving optimality and detecting suboptimality in research products traded on the knowledge exchange market. Concerted action is needed to face these challenges.

---

[5] Many of these suggested changes are relatively standard in other fields like psychology. We argue that entrepreneurship scholarship could catch up, which should be feasible.
[6] The economic literature also uses the term 'peaches' (Kreps, 1990) or 'creampuffs' (Bond, 1982) for high-quality products, in contrast to 'lemons'.

## 2. AI Revolution Well Under Way

Despite the hype around AI, we recognize that many scholars in our field may be somewhat hesitant or even critical of the actual scope of AI adoption and its disruptive potential for the field (Lévesque et al., 2022). Before we explore the market-for-lemons dynamics that we see unfolding, we therefore address the fundamental question of whether deeper consideration of the effects of (potentially) massive AI adoption in the field—the starting point of this article—is premature at this stage and based on mere speculation about massive adoption. As we demonstrate next, we believe it is not since the adoption is already in progress.

Advanced AI methods are one of the most powerful and disruptive forces in the history of humankind (Suleyman, 2023). AI's transformative potential has been described in multiple ways, including as having the potential to create societal transformation levels similar to former general-purpose technologies (GPTs) like electricity, a transformation analogous to the industrial revolution, and as having a metamorphic impact on society (Gruetzemacher & Whittlestone, 2019). Academics, AI and IT experts, government policy and law makers, technology and business groups, the news media, and ethicists, among others, are engaged in heated debate on the great tension between the massive positive potential of AI (e.g., by serving humans and their humane societies) and the existential risks where "creating a superintelligence misaligned with human values could lead to catastrophic outcomes" (Jones, 2023a:0).

The discussions on machine learning, deep learning, and the many applications that amaze so many, particularly with the recent rise of large language models and related generative AI tools (e.g., ChatGPT; Chow, 2023), have increasingly intensified. AI applications continue to impress with some great performance. For instance, AI helps doctors to better diagnose clinical diseases and find effective therapies and drugs (Hasselgren & Oprea, 2024; He et al., 2023; Klauschen et al., 2024). Such supportive effects are also widely discussed in academia, including discussions on a major paradigm shift in academic writing (e.g., AI as a co-author) (Dwivedi et al., 2023; Flanagin et al., 2023; Golan et al., 2023). Concurrently, societies' most eminent AI scholars are forming coalitions to express urgent concerns regarding massive risks (e.g., biases, misuse) associated with advanced AI methods (e.g., Bengio et al., 2023). Similarly, prominent AI entrepreneurs depict a profoundly perilous vision of AI's future capabilities and its impact on humanity (CNN, 2023).

In view of this unfolding technological breakthrough, and the resulting tensions between positive opportunities and existential threats, editors of top business journals, along with those in many other disciplines, have placed transformative AI in center stage (Grimes et al., 2023;

Haenlein et al., 2022; von Krogh et al., 2023), including the discussion and presentation of new AI policies for these journals (e.g., Gatrell et al., 2024). A wave of new special issues on the topic has also emerged across various journals (e.g., Berente et al., 2021; Fosso Wamba et al. 2022; Jia et al., 2024; Obschonka & Audretsch, 2020; Obschonka et al., 2024; Raisch et al., 2024). Management scholars have been engaging in debates on how to govern AI in organizations (Bankins & Formosa, 2023; Goos & Savona, 2024), and not just to create and capture value (e.g., Akter et al., 2022; Berg et al., 2023; Moser et al., 2023, Shepherd & Majchrzak, 2022), but also to prevent its detrimental effects (e.g., cyber-threats, Bécue et al., 2021; unprecedented ethical issues, Landers & Behrend, 2022). Entrepreneurship scholars have started to investigate how AI can support not only entrepreneurs (Chalmers et al., 2021; Otis et al., 2023; e.g., via prompt engineering for generative AI; Short & Short, 2023), but also other important stakeholders in the entrepreneurial process (e.g., see Blohm et al., 2022, for investors; Hannigan et al., 2022, for policymakers; or Shepherd & Majchrzak, 2022, for educators). Moreover, new questions such as how to best leverage scarce AI resources—from AI expert knowledge or access to big data and supercomputers or cloud computing—to promote entrepreneurial activities have gained significant momentum (Gofman & Jin, 2024; Obschonka & Audretsch, 2020). Finally, entrepreneurship practice is of course at the forefront of the AI revolution, exemplified by a myriad of AI startups.

How far will this revolution lead us? It is of course difficult to predict the future, but we might see more disruption than we think. To illustrate, a common slogan in our field is "entrepreneurship requires action" (McMullen & Shepherd, 2006:132) to navigate through the infamous 'uncertainty fog' faced by founders of new businesses. In yesterday's logic of the entrepreneurship field, entrepreneurial action was intimately intertwined with the human agency of entrepreneurs. Startups did not develop and thrive out of thin air, the entrepreneurs drove the processes, with more (or less) control (i.e., entrepreneurship as a human artifact, Sarasvathy, 2023). The human-entrepreneurship nexus is foundational in our understanding of entrepreneurship and its drivers (Shane, 2003). However, yesterday's logic may no longer apply at some point in the future because entrepreneurship *might* involve human entrepreneurs (e.g., for legal or ethical reasons), but it might not necessarily *require* their involvement. Importantly, akin to Hal in the 1968 epic science fiction movie '2001: A space Odyssey,' computers might acquire entrepreneurial agency by interreacting and learning from each other in relatively uncontrolled and unsupervised ways without much guidance or control from humans (an old and fascinating debate in the field of computer science; Bostrom, 2014; Heimans & Timms, 2024).

Hence, in comparison to a fast-advancing web of machine intelligence, agentic humans with limited intelligence compared to AI might get overthrown from the entrepreneurship command tower. This is of course not a new discussion, but part of the larger discussion in society on whether intelligent machines will replace humans for conducting important and/or difficult tasks (e.g., Chui et al., 2016). Entrepreneurship is a difficult task primarily due to its inherent uncertainty, at least for humans, but this may differ if AI could find it easier (Obschonka & Fisch, 2022). If uncertainty is core to the entrepreneurship phenomenon, and if the AI revolution eventually and ostensibly eliminates this uncertainty, then we must conclude that the research field might implode (e.g., because the field would predominantly focus on relatively predictable management matters; Sarasvathy, 2001). However, it is unlikely that we will reach this point soon, and maybe it will never happen. Nonetheless, to help us to be prepared, we next share our thoughts on the potential unfolding of a market for lemons in entrepreneurial studies, and what interventions we can initiate if this occurs, given that the AI revolution is upon us.

## 3. The Market-for-Lemons Concept

Social science scholars, particularly those with economics or business backgrounds, may be familiar with George Akerlof's (1970) seminal work on 'markets for lemons', and his highly influential investigation of product quality uncertainty and the resulting market mechanisms that unfold. In his example, the used car market, he distinguishes between good and bad cars, the latter of "which in America are known as 'lemons'" (p.489). In his classic 'lemon' metaphor, bad cars are sold in this used car market—whereby the buyers discover hidden defects or problems *after* they buy them. Akerlof contends that car purchasers (buyers) do not possess the information that used-car dealerships (sellers) have (e.g., incipient problems in the engine or transmission[7]) to differentiate a good car from a defective one—a lemon—which could eventually result in a market where only the lemons are sold by used-car dealerships. This is because, due to information asymmetry, buyers are unable to distinguish between good cars and bad cars. As a result, they are only willing to pay an average price that reflects the risk of getting a 'lemon', which can drive good cars out of this ineffective market. With such a market failure, the whole used-car industry is negatively affected. Hundreds, if not thousands, of studies have since shown that 'lemons' are not only restricted to used cars but extend to myriad products affected by information asymmetries in their respective market (e.g., insurance products, Downing et al., 2009; pharmaceuticals, Light & Lexchin, 2021). This has impressively demonstrated that the market-for-lemons concept deserves careful consideration across a wide range of products and their markets—a principle we adhere to in this article.

[7] These problems would cost several thousand dollars to repair but cannot be identified by the buyer, yet the seller may know about the problem, likely from their knowledgeable mechanic, which could prompt them to sell their 'lemon' to an unsuspecting buyer.

To this end, let us delve deeper into the more formal theorizing of Akerlof's concept. His own conclusions were derived by considering a market for used cars that contains a proportion $\alpha$ of bad cars (lemons) while the remaining proportion, $1-\alpha$, are good cars. Buyers are willing to pay up to $p^B$ for a bad car and up to $p^G$ for a good car. On the seller side, used-car dealerships value a bad car at $v^B$, which does not exceed what a buyer is willing to pay for a bad car (i.e., $v^B < p^B$). Similarly, used-car dealerships value a good car at $v^G$, which does not exceed what a buyer is willing to pay for a good car (i.e., $v^G < p^G$). In his seminal 1970 article, Akerlof argues that if no market regulations exist and the buyers are unable to observe the real quality of the used cars (whether it is optimal or suboptimal), then the bad and the good car markets would unite to form only one market and sell all used cars at the same price. An 'average buyer' would then be willing to pay at most $\bar{p} = \alpha p^B + [1-\alpha]p^G$. Consequently, if the sellers' value $v^G$ for a good car exceeds what the 'average buyer' is willing to pay (i.e., $v^G > \bar{p}$), then used-car dealerships of good cars will eventually exit the used-car market and leave us with bad used cars—the lemons.

Akerlof's (1970) seminal theory of asymmetric information[8] is viewed as one of the most important contributions to the economics of information (Löfgren et al., 2002) and helped Akerlof, along with his colleagues Michael Spence and Joseph Stiglitz, win the 2001 Nobel Prize in Economics. Of course, there is more to this theory that is relevant to our arguments. For instance, car dealerships with good used cars can take actions to counteract information asymmetry that creates a market for lemons—for instance, by signaling to buyers that their used cars are indeed good. This countermeasure extends the theory of asymmetric information to the theory of signaling. Signaling can be particularly important if car buyers can be protected by institutions (e.g., professional dealers' guarantees) to avoid the 'adverse selection' problem whereby, due to buyers' lack of information on used cars' quality, dealerships of bad used cars may force out everyone else from their market side, thereby eliminating advantageous transactions for both sides. Akerlof has used his theory and its key derivatives—signaling and adverse selection—in a variety of contexts.[9]

[8] Essentially, asymmetric information theory suggests that sellers often possess more information than buyers, thus skewing the price of goods sold.
[9] For instance, in sharecropping contracts between landlords and tenants, where the information asymmetry hurdle could be mitigated by less-informed landlords paying tenants (would be 'lemons') a fixed share from the harvest to generate greater efforts from these tenants (Akerlof, 1976).

Our main argument is that Akerlof's (1970) theory of asymmetric information and the resulting market for lemons could also apply in the entrepreneurship field with its current standards and regulations or, more precisely, lack thereof. Noteworthy is that we are not the first to use the market-for-lemons analogy in the academic publication context. Cottrell (2013) uses the market-for-lemons framing to discuss academic integrity (though not specific to AI nor entrepreneurship). As our article unfolds, we will detail three key differentiators. First, we consider different sellers and buyers, who are represented in Cottrell (2013) by, respectively, the publishers and paying readers; instead, we focus on authors and readers that include reviewers and editors. Second, the mechanisms that create information asymmetry differ. In Cottrell (2013), the publisher possesses more information than the paying readers on an article's quality, since the publisher benefits from all information gathered through the review process; instead, we offer two types of information asymmetry that involves the use of (advanced) AI methods, authors, and readers. A third key differentiator pertains to the consideration of non-monetary exchanges in a market for scientific publication; we consider new knowledge exchanges.

Hence, for productive knowledge exchanges, the field's *status quo* (with its current standards and regulations), which is healthy in terms of yesterday's logic—that is, in the pre-AI revolution—might become increasingly unviable. Importantly, the *status quo* might not only help establish a market for lemons, but also create reinforcing mechanisms that maintain and can amplify such an undesirable market. Therefore, we put forth that the entrepreneurship field is particularly prone to morphing into a market for lemons due it its deep 'DNA'. We explain this claim first by exploring the entrepreneurship field's roots and evolution, and highlighting how its evolution leads to present-day triggers of market-for-lemons dynamics, fueled by the unfolding AI revolution. We highlight a double-black-box puzzle and then proceed with concrete examples of how the market for lemons could develop in our field, and what guidelines could prevent it.

## 4. Entrepreneurship Scholarship Evolution: Market-for-Lemons Roots?

From time to time, research fields undergo transformational phases. Entrepreneurship scholarship, as a major research discipline and community, evolved "from a small emerging venture in the 1970s to a global industry today with thousands of people around the world who consider themselves entrepreneurship scholars" (Landström, 2020:1). Many scholars agree that this evolution did not take a simple linear path but occurred over multiple transformations during various phases. Following Landström (2020), these transformations occurred, as summarized in Figure 1, during the field's transitions from (1) seminal early theorizing and empirical contributions in mainstream disciplines (e.g., economics, management, innovation, psychology, political science, sociology, finance, and geography), to (2) the formation of the actual research field of entrepreneurship in the 1980s, to (3) the growth of entrepreneurship research in the 1990s, to (4) the conceptual recalibration and maturation of entrepreneurship research during its golden era[10] in the early 2000s when the field 'exploded' by anchoring its scholarly definitions and orientations, to (5) the 2010s when the exploding field transitioned from a "'low entry' field resulting in a 'transient' research community" in the 2000s (Landström, 2005:67) to a 'high entry' field (with respect to publishing in the field's top journals), with a more stable research community, thus critically contributing to the established high-status of the field today.

Figure 1. Entrepreneurship scholarship evolution

| Early theorizing, empirical contributions, and forerunners of mainstream disciplines | Formation of the actual entrepreneur-ship research field | Initial growth | **Golden era:** conceptual recalibration and maturation | Transition from low-to high-entry field | **AI revolution begins*** | **What Does the Future Hold for Entrepreneurship Scholarship?** |
|---|---|---|---|---|---|---|
| pre-1970s | 1980s | 1990s | 2000s | 2010s | | 2020s |

*Note.* The evolutionary pathway prior to the AI revolution is based on Landström (2020).
* Around 2020 (Chalmers et al., 2021; Obschonka & Audretsch, 2020; Shepherd & Majchrzak, 2022).

Within its more mature stage today, the field has branched out not only in terms of growing subfields such as social entrepreneurship, sustainable entrepreneurship, entrepreneurs' mental health and well-being, entrepreneurial finance, entrepreneurship education, and so on, but also in terms of its growing impact on other academic fields like psychology, health, genetics, culture, industrial organization, macro-economics, and public policy (Landström, 2020; Thurik et al., 2023). This branching out can be seen as another transformation of the field. But what triggered these historical transformations and what can be learned for the unfolding AI transformation of the field? Some scholars attribute the historical transformations to massive *external* changes that pushed entrepreneurship research into these transitions and associated growth. A key driver was technological progress (e.g., in information and communication technologies). Another important driver could have been the 'triumph' of capitalism over communism in the 1990s (Rona-Tas, 1994; Frese et al., 1996) and the subsequent rise of entrepreneurship in developing countries such as China (Tsang, 1996), underpinned by globalization. Similarly, the accelerated transition from managed economies to knowledge-based and thus innovation-focused economies also played a role (Audretsch, 2009), leading to policymakers acknowledging that “the key to economic growth and productivity improvements lies in the entrepreneurial capacity of an economy” (Prodi, 2002). Hence, economic and political changes were major drivers that ushered in the importance of entrepreneurship, which soon became a buzzword in the elite circles of society (Shane, 2008), and then permeated into mainstream society and public media as, for instance, entrepreneurs became the subject of acclaimed blockbuster movies and bestselling books (Isaacson, 2011, 2023).

[10] Landström (2020) coined this era as the field's *Golden Era*. While we think that this might underestimate the positive developments that occurred after the 2000 decade in the field, we chose to use this expression because it stimulated our thinking around the positive impact that AI can have on our field, moving toward a new golden era.

External triggers linked to technological progress and economic, cultural, and political changes are again also driving the ongoing AI transformation of the entrepreneurship field. Clearly, the primary external trigger for this transformation lies in the massive technological advancements and progress in AI and big data, and their concrete availability to researchers. We have come a long way from Alan Turing’s visionary questions such as “Can machines think?” (Turing, 1950:433) to Moore’s law of exponential growth in computing power (Lundstrom, 2003) confirmed by present-day supercomputers and powerful AI methods interacting with unprecedented amounts of data—all of which are becoming increasingly difficult for humans, including researchers and scientists, to comprehend and control quality as they employ such AI methods in their studies (Messeri & Crockett, 2024).

In contrast to external triggers, more hidden *internal* triggers exist that drove historical transformations of the entrepreneurship field, and importantly, also channel the new transformation in certain directions. Historically, the rise of entrepreneurship as a phenomenon was also driven by internal triggers (push factors) stemming from the heart of entrepreneurship research itself. For instance, scholars empirically confirmed that productive entrepreneurship is a driver of industrial evolution, job creation, innovation, and growth (Birch, 1987; Davis & Haltiwanger, 1992; Klepper, 1996; see also Guzman & Stern, 2020). Moreover, empirical evidence on the psychological utility of entrepreneurial work as a viable career choice option (Benz & Frey, 2008; Obschonka et al., 2023; Van Praag & Versloot, 2007) in an era of growing recognition of the valuable non-financial payoffs from work such as purpose, meaning, and eudaimonic well-being (vs. the pursuit of pleasure and enjoyment) (Dik et al., 2015; Ryff, 2019), have contributed to this rise. In the next section, we outline the specific internal triggers that we believe are shaping the current AI transformation of the field—potentially steering it toward a market-for-lemons scenario if left unaddressed.

## 5. Today's Entrepreneurship Scholarship: Internal Triggers of Market-for-Lemons Vulnerability

Acknowledging the impact of technological progress in AI and big data, along with related economic, cultural, and political changes (as the external triggers we discussed above), we want to focus on *internal* triggers of a potential AI-driven transformation in the entrepreneurship field that could lead us toward an unfolding market for lemons within the research landscape. Unlike the external triggers, these internal triggers originate deep within the 'DNA' of the research field—products of its unique evolution. Hence, they are somewhat hidden, yet important to uncover. In the tradition of the phenomenon orientation of the field, we must bring a crucial aspect of entrepreneurship as a phenomenon: *uncertainty*. The essence of entrepreneurship lies in understanding and effectively managing uncertainty to drive innovation and create value (Knight, 1921; see also Alvarez & Barney, 2005; McMullen & Shepherd, 2006). Entrepreneurship is also a 'mutation' of adjacent disciplines and fields such as management, psychology, sociology, economics, and/or geography, though its fundamental difference lies in the nature of the uncertainty in the entrepreneurial process. At the dawn of the massive AI revolution, the uncertainty aspect of entrepreneurship, combined with that revolution's unknowns, leads us to highlight six (internal) triggers of today's entrepreneurship scholarship for creating a particular vulnerability to a market for lemons.

*Construct validity*. Though accompanied by some warnings (e.g., Roberts et al., 2024), qualitative analysis, for instance, has largely benefited from AI methods that can help analyze vast amounts of information to refine and validate constructs (e.g., large language models, which are trained on unstructured data, are great at finding patterns in a sea of noise). In entrepreneurship, however, the difficulty of clearly defining key constructs based on a broad consensus and empirical precision suggests this first trigger (Davidsson, 2015). Construct validity is among the most important quality indicator of a research field (Cronbach & Meehl, 1955; Peter, 1981). Yet, defining and measuring a construct like 'entrepreneurial opportunity' is difficult as the construct is nebulous. Hence, developing and empirically testing a coherent body of theoretical and empirical knowledge around such a construct, which in turn can guide practice (Davidsson, 2015), becomes nearly impossible. Arguably, the well-known recent 'opportunity dispute' is a concrete example of this struggle for concept clarity, as it has shaken the field's foundations. This dispute stimulated important questions such as "What is an entrepreneurial opportunity?" and "How can we define and, thus, research it and predict its occurrence in the real world?" (Alvarez & Barney, 2020; Davidsson, 2015; Foss & Klein, 2020). Another difficulty emerges from diverging views on measurement coherence (Tal, 2013). Clark et al. (2024) provides an example with entrepreneurial orientation (EO) being an individual-level construct exclusively, viewing the application of EO at a higher level of analysis (e.g., firm level) as bad science. Moreover, the entrepreneurship phenomenon is a moving target subject to unpredictable change and transformation that yield ever-changing dynamics in which entrepreneurship is embedded (Davidsson et al., 2020), thus complicating the construct measurement formula.

*Theory building*. AI methods also can process and synthesize thousands of pages to facilitate theory building (e.g., as of June 2024, Gemini 1.5 Pro had two million tokens context window). Nonetheless, entrepreneurship scholars have highlighted the challenge in major attempts to build unique and novel entrepreneurship theories (Chrisman et al., 2023; Shepherd & Williams, 2023; Zahra et al., 2023). This challenge arises from complications when empirically testing theories in a Popperian sense[11]—and this often leads theorists to describe their theoretical attempts as 'frameworks' or 'entrepreneurial methods', instead of predictive theories (Davidsson et al., 2020; Sarasvathy, 2023; see also Grégoire & Cherchem, 2020). Hence, while searching for more distinct theories that can be directly tested (e.g., falsified empirically), the field has leaned heavily on theories from other disciplines and their applications in the entrepreneurship context (Zahra et al., 2023). Particularly creative, comprehensive, and laudable theorizing attempts from within the field (e.g., Davidsson et al., 2020; Sarasvathy, 2023; McMullen, 2023) have also received criticisms in the entrepreneurship literature (e.g., Arend, 2020, 2023; Arend et al., 2015; Alvarez & Barney, 2020; Wood, 2017). We may simply conclude, at least for now, that it is difficult to put the multifaceted construct of entrepreneurship into a scientific formula. Therefore, we must recognize that writing about theory is a challenging and complex topic (Cornelissen, 2017), especially theory about an interdisciplinary phenomena like entrepreneurship (see also Chrisman et al., 2023; Zahra et al., 2023); however, in the words of Lewin (1952:169), "nothing is more practical than a good theory." Prioritizing domain-specific theory-building (plus testing it) is thus necessary for the future of our field.

*Research relevance*. Relevance in research means addressing key research questions that (1) inform practitioners, policymakers, and scholars on current issues, along with justifiable assumptions and hypotheses (or propositions), (2) offer rationales that apply directly to their work, and (3) provide research findings and implications that clearly inform them (Toffel,

[11] Testing theories empirically, subjecting them to rigorous testing and potentially disproving or proving them through observation or experimentation.

2016). AI's capacity to have access to the cumulative knowledge of humanity while simultaneously accessing current information can help identify emerging trends and research gaps that can potentially fulfil the above requirements for research relevance. However, AI might further separate scholars from the context they hope to study. In other words, scholars who will become obsolete are those who do not understand the context of entrepreneurship, since they will certainly be less efficient than AI. This pertains not only to interacting with entrepreneurs as they design studies but also to how they approach analyzing data. AI methods could exponentiate this issue if entrepreneurship scholars are not ready to 'get their hands dirty.' Moreover, too often the 'outside world' may assess the value of entrepreneurship research not so much in terms of theoretical contributions but in terms of concrete, practical implications of related scientific evidence, even though "theory enriches managerial practice" (Zahra et al., 2023:3). With the triggers described above, the field requires scholars to dedicate extra effort, creativity, and engagement to show relevance.

*Publication pressure*. Publication pressure remains a common feature in our field, with scholars often required to produce work for promotion or to meet key performance indicators (KPIs). This pressure can fuel market-for-lemons dynamics due to the large number of research articles, which are now increasingly influenced by AI. The automation of routine tasks facilitated by AI such as literature review, instrument design, cleaning and analyzing data, describing results, and copy editing, to name a few, can speed up the writing of scholarly articles to attenuate the publish-or-perish academic environment (Samuels-White, 2024). In other words, entrepreneurship scholars could face increased productivity and efficiency from using AI, but the increased use of AI could also make it more challenging to accurately assess research quality due to information asymmetries and, thus, enlarge the size of a potential market for lemons in the field. Moreover, the impact of AI-driven scholarly research can be questioned and create other kinds of publication pressure. An example is the major definitory advances of Shane and Venkataraman's (2000) epochal article on the entrepreneur-opportunity nexus, which was initially enthusiastically celebrated as the entrepreneurship field's unifying scientific guidepost, but since then has faced considerable criticism about its actual constructive impact on the field (Alvarez & Barney, 2020; Davidsson, 2015; Foss & Klein, 2020), with some scholars explicitly questioning such foundational paradigms that were long thought to be the rock on which modern entrepreneurship scholarship stands (e.g., Davidsson, 2023). Other scholars are diverging to philosophical approaches (Ramoglou & McMullen, 2024) or religious faith (Pidduck et al., 2024) to create a new conceptual orientation. Also noteworthy is that during the field's prolific golden era of the 2000s, major articles generating thousands of citations could claim uncharted academic territory relatively easily as first movers. However, many pertinent topics are relatively well researched today, leaving less uncharted territory for potential influential academic endeavors and literature-gap filling. Concurrently, the scientific bar is much higher today for making major theoretical and empirical contributions that are truly new and innovative because the complexity of today's world requires advanced theorizing and highly complicated empirical research design. Nevertheless, this does not mean that the field is shrinking and that scholars are giving up—rather the opposite is true—but scholars have come to question what constitutes truly strong, novel contributions. Entrepreneurship scholars increasingly lament about 'too much noise' in the literature, referring to hundreds of new, but not necessarily impactful, published articles each month in an increasingly larger number of entrepreneurship journals.[12]

*Education and Training*. With the maturity and size of the field, and its *bona fide* status, also come new responsibilities in terms of educating and training the next generations of entrepreneurship scholars. Other than in the 2000s' golden era, where many (if not most) entrepreneurship scholars entered the field from other social science disciplines with their respective theoretical and empirical training, the entrepreneurship field is now educating and training its own talent. New generations of entrepreneurship scholars are thus socialized primarily using the current entrepreneurship literature, be it in terms of ongoing conversations in that literature or as gap spotting for the application of theories and constructs from outside the entrepreneurship field. Hence, whatever happens to this literature and the field also directly affects the academic education and training of the most important agents in our field—the next generation of entrepreneurship scholars. We also recognize that AI can provide personalized education and potentially solve Bloom's (1984) '2 sigma problem', whereby students educated under a one-to-one tutoring approach using mastery-based learning were shown to achieve, on average, two standard deviations better than those in the classroom. Nonetheless, careful consideration associated with education and training—and particularly on the proper applications of advanced AI methods—is warranted for the future well-being of our field.

*Inter-, trans-, and multi-disciplinarity*. The sixth trigger is the interdisciplinary, transdisciplinary, and multidisciplinary nature of entrepreneurship research. Davidsson (2023) contend that the field has evolved not only into a multidisciplinary, but also multi-level and multi-methodological field. While this can be seen as a strength for the field's vitality and creativity, it also introduces additional complexities, such as integrating diverse perspectives and methodologies (Baker & Welter, 2018). Moreover, entrepreneurship scholars are confronted with myriad influencing factors that require investigation at various levels (e.g., individual, firm, and context) from various domains (e.g., human and social capital, finance, economics, innovation, psychology, networks, teams, culture, and policy), necessitating multidisciplinary approaches and teamwork. This collaboration is needed for theorizing and empirical development but also for practical reasons. For instance, diverse research teams must speak the same academic language even though they are trained in different disciplines, theoretical paradigms, and/or terminologies, and they require clear communication and understanding, without which the research process can faulter or fail (e.g., if research involves AI experts and non-AI experts unfamiliar with AI; Galsgaard et al., 2022).

[12] A business scholar at a major U.S. university recently told us about a conscious exaggeration with a wink that "there are almost more entrepreneurship scholars nowadays than entrepreneurs," pointing to the increasing crowdedness of the academic literature and thus making it more difficult for readers outside the field to identify and concentrate on particularly impactful contributions.

As they relate to the black-box characteristic of the entrepreneurship phenomenon, the six triggers of today's entrepreneurship scholarship bring concerns regarding our field's potential incapacity to detect lemons. In Figure 2, we show this black box intertwined with these triggers and call it BLACK BOX 1. We foresee the powerful impact of another, well-known black box in the presence of advanced AI methods in research processes (Klauschen et al., 2024; Messeri & Crockett, 2024; Rai, 2020) owing to the opacity (i.e., difficult to understand) from using these methods and evaluating their precision—we call it BLACK BOX 2. We portray BLACK BOX 2 as a *flashlight* to 'brighten'—thus amplify—the impact of the entrepreneurship scholarship's triggers when using AI methods due to these methods' opacity. Hence, both black boxes combined lead to a *double-black-box puzzle* of entrepreneurship scholarship, further exacerbating the potential for a market for lemons in our field.

Let us illustrate how this double-black-box puzzle interacts with the six triggers, while recognizing that these triggers can also act *collectively* to amplify the market for lemons. We begin with *construct validity*. If we consider medical research, for instance, the predictive tasks of a machine learning model can be relatively clear owing to readily defined constructs like *clinical disease* (e.g., the ICD-11 manual of the WHO; World Health Organization, 2022), which then becomes the target of therapeutic research and intervention (Hasselgren & Oprea, 2024). In entrepreneurship research, however, construct validity is more difficult to determine (e.g., an entrepreneurial opportunity is a nebulous concept) and the use of advanced AI methods can create additional opacity when it learns relationships and patterns from data rather than precise programming, which creates additional unknowns that make evaluating the research arduous. While the entrepreneurship field may continue to debate some if its key constructs and their definitions (Alvarez & Barney, 2020; Davidsson, 2015; Foss & Klein, 2020), advanced AI methods are successful at delivering discoveries with often substantial precision[13] in fields that have a greater consensus on constructs and definitions. Thus, in the near term, we could experience some difficulties owing to such debates, but in the longer term, AI application could resolve these debates.

Figure 2. The double-black-box puzzle of entrepreneurship scholarship

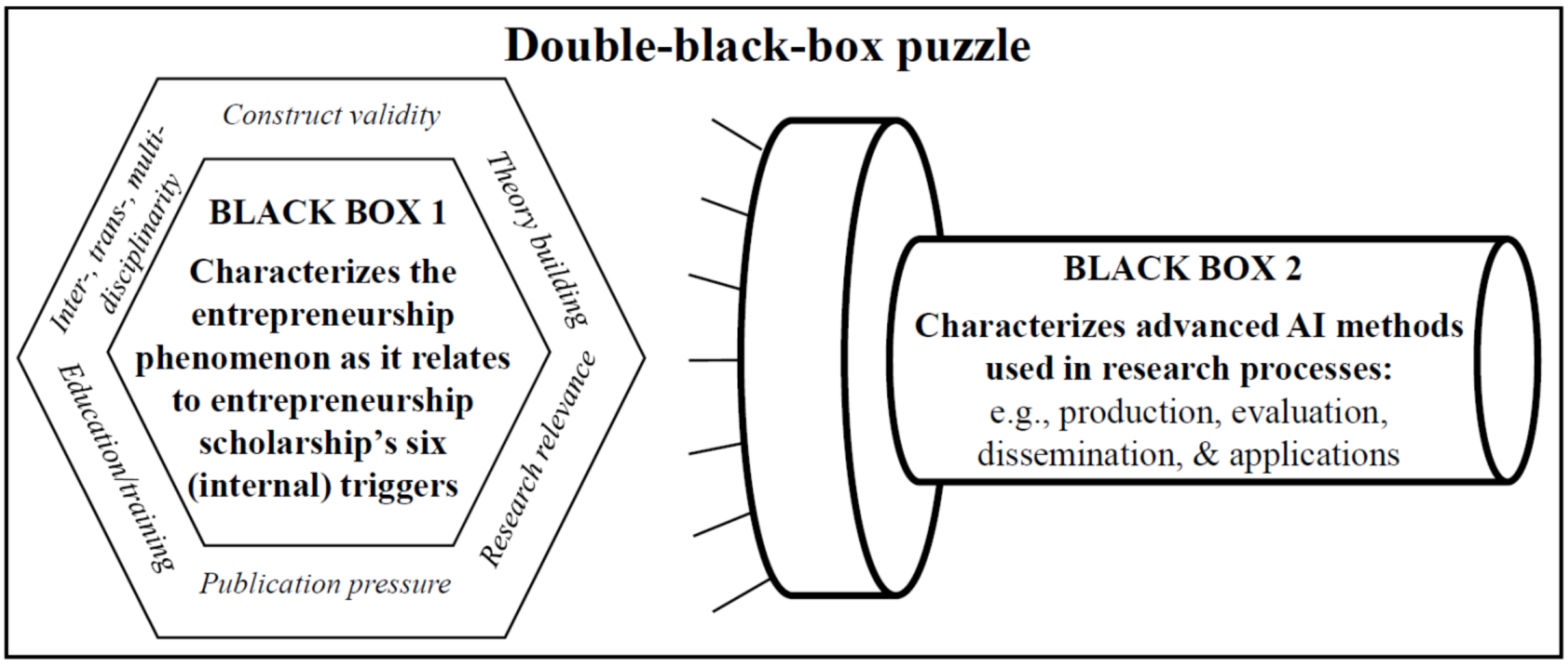

The second trigger in entrepreneurship scholarship—*theory building*—also amplifies the impact from the opacity of AI application. Through 'super exploration' from applying AI to big data, researchers might be able to discover predictive patterns that help them to build new theory that *human* intelligence alone would not have been able to discover (Obschonka & Fisch, 2022), since advanced AI methods can indeed be used in strategic ways to build (and test) theory (e.g., Chou et al., 2023; Lévesque et al., 2022). While the new theory would be based on the predictive capacity of the AI method, the opacity of the method could render it quite difficult, if not impossible, for authors and readers (e.g., reviewers and editors) alike in the entrepreneurship field to fully grasp the logical process of how this theory was developed, because they are typically not trained to understand the 'mechanics of the engine under the [AI] hood'. Nonetheless, in the longer term, education and training are likely to address this issue.

The four remaining triggers—*publication pressure*, *research relevance*, *education and training*, and *inter-, trans-, and multi-disciplinarity*—can also amplify the impact from the opacity of AI method in their own unique way. Publication pressures may encourage more 'lemons' within the field owing to an ever-increasing volume of difficult-to-detect suboptimality in research products using AI methods. Additionally, in a world of emerging AI-driven abundance (Grimes et al., 2023), we might see a big leap forward in research productivity, potentially further increasing the size of the market for lemons. With so many lemons, research relevance (thus impact) might suffer as well—not just because of the sheer number of new research insights but because it might possibly take years for the lemons to be debunked as flawed products via application in practice. We could also face difficulty training our own talent in the field by using flawed research insights from within the field. Basing education and training on flawed research would have a negative long-term effect on the field's most precious resource, humans, assuming AI will never replace human researchers. In other words, the undetected intergenerational transmission of lemons could perpetuate the market for lemons.

[13] Yet, the potential for bias (e.g., AI hallucinations or social bias and discrimination) must be considered.

Lastly, entrepreneurship research often involves a combination of disciplines (e.g., management, psychology, sociology, economics, statistics, law). The resulting complexity of this inter/multi/trans-disciplinary research process and its products could further contribute to the information asymmetries between authors and readers, and to a greater degree than in fields like sociology or psychology where AI revolution discussions (e.g., respectively, Brandt & Timmermans, 2021; Vezzoli & Zogmaister, 2023) have already taken place. Overall, relative to other disciplines, the contemporary landscape of entrepreneurship research, as a product of its unique evolution and, consequently, foreseen internal triggers of today's entrepreneurship scholarship, exposes the field to certain vulnerabilities for market-for-lemon dynamics associated with massive AI adoption in research.

## 6. Entrepreneurship Scholarship's Unfolding Market for Lemons: Dominant Knowledge Asymmetries

Whereas the mission of scholarly fields involves guarding existing knowledge, the challenging core business is the production and dissemination of *new* knowledge. Typically, the most influential knowledge production activity, scholarly articles, appear in the field's top academic journals. These newly accepted manuscripts (including peer reviewed articles and editorials) represent the endpoint of comprehensive research processes and a starting point for impact (e.g., in academia, practice, policy, and education). Impact is linked to knowledge dissemination—how new knowledge is then applied in academia and practice, and with what consequences (e.g., useful, useless, or detrimental outcomes or applications). To be *productive* new knowledge, the knowledge exchanged on these markets must be of high quality and flawless, and the market processes, such as the transactions between sellers' offerings and buyers' purchases of this new knowledge, must be efficient and unimpaired. Only then do these markets truly serve their stakeholders, particularly entrepreneurship academics. Nonetheless, such markets can be fragile or falter under certain circumstances and during certain transitions. This, in turn, requires careful attention and consideration, as exchanges on these markets involves the core business driving the vitality, reputation, and impact of scholarly fields. In this spirit, following Akerlof's (1970) logic, we see two types of information asymmetries to next argue that these healthy market exchanges could become disrupted, particularly in the near term as the AI revolution evolves but effective standards and regulations remains unclear and thus missing.

The first type of information asymmetries can occur between the advanced AI methods and underlying data, on one side, and the scholar(s) employing these methods and data for their research (empirical or theoretical), on the other side.[14] AI-based research is known to pose substantial opacity challenges even to scholars familiar with AI methods, due to their black box nature (Klauschen et al., 2024; Messeri & Crockett, 2024; Rai, 2020), as portrayed in Figure 2. Indeed, a significant difference exists between employing a regression analysis on a sample of, say, 100 study subjects, versus using machine or deep learning to analyze large volumes of data, or consulting large language models. While an author has many options to check the validity of regression analyses and can even examine the data personally to validate patterns, advanced AI methods make validation much more challenging. This difficulty extends beyond the comprehensibility of AI results to concerns about deeply hidden biases and other limitations (Landers & Behrend, 2022). Consequently, it is fair to believe (at least in the near future) that some scholars might apply these AI methods without fully grasping the nuances, limitations, or appropriate contexts for their use, which *could* result in suboptimal research outputs that go undetected (even by readers like reviewers trusting the authors' evaluation).[15] In other words, if the methods themselves are so complex that even the users (the authors) cannot fully understand or validate them, this creates a gap—an asymmetry—where the methods 'know' more than the users about the validity of the results.

---

[14] For instance, using deep learning to identify patterns in a dataset that predict future outcomes, or employing large language models to help generate research questions or develop new theory.

[15] In fact, even with regression analysis where the researcher upload the data, choose a model, click a button, and the software spills out the output, for many scholars it is a black box. They know the application of the method and how to interpret the results, though often at the expense of checking whether the underlying statistical assumptions of regression analysis hold. For instance, logistic regression assumes absence of multicollinearity among the predictors, linearity between the independent variables and log odds of the dependent variable, independence of errors, and lack of strongly influential outliers. Bayesian additive regression tree (BART) models (Chipman et al., 2010), on the other hand, do not need to satisfy these statistical assumptions and they can also capture high-order interactions along with the ability to interpret the results. Although using AI methods may save a few extra steps and make it easier for scholars to use more complex methods (to avoid the assumption-violation danger), the deeply hidden biases and other limitations remain.

The second type of information asymmetries is between authors and readers (e.g., peer reviewers but also editors) and we consider six contributing mechanisms. The first mechanism is *technical expertise* whereby authors using AI methods may possess specialized technical knowhow and expertise that go beyond what is common among general readers or reviewers, including a better understanding of the algorithms, data processing techniques, and the specific ways AI is applied in their research. The second mechanism is the *complexity of AI methods* whereby particularly advanced ones are often complex and involve intricate models and large datasets, which can make it difficult for readers and reviewers without a background in AI to fully grasp the nuances and implications of the research, creating a knowledge gap. The third mechanism is *implicit knowledge* whereby much of the understanding of AI methods comes from knowledge gained through hands-on experience and experimentation. Implicit knowledge is typically more difficult to fully convey to others (compared to explicit knowledge) in a scholarly manuscript, leading to a situation where authors have a better, more intuitive understanding of their methods compared to reviewers and readers.[16] The fourth mechanism is the *AI's evolution* whereby new techniques and methodologies are continuously emerging. Authors actively engaged in AI research are likely to be more up to date with the latest advancements, while reviewers and readers might not be as current, especially if their primary focus diverges from using AI methods. The fifth mechanism is *communication challenge* whereby AI-based research often involve complex data and outcomes difficult to communicate clearly in written form, particularly in traditional concise and short article formats. Even well-explained studies might not fully bridge the gap in understanding for those not deeply familiar with the AI methods used, leading to potential misinterpretation or underappreciation of the research's significance. The final mechanism is *interdisciplinary barriers* whereby AI-based research often intersects with various other disciplines, which can further complicate communication. Reviewers and readers might have expertise in the domain where AI is applied but not in the AI methods themselves, exacerbating the information asymmetry.

Importantly, both types of information asymmetries and allied mechanisms could be further compounded if reviewers use generative AI tools "to summarize, critique, and evaluate manuscripts", creating "additional challenges for journals in avoiding both 'false negatives,' wherein potentially groundbreaking work is rejected, as well as 'false positives,' wherein inaccurate, biased, or even fabricated studies are accepted" (Grimes et al., 2023:1620). Hence, if information asymmetries do play out, this could drive the market for lemons. In this market, authors' and reviewers' capacity for detecting suboptimal work, the lemons, could be limited. In other words, the two types of information asymmetries will prevent some scholars from knowing with certainty whether some research products are suboptimal works. Yet, the mere fact that suboptimal research products *could* circulate undetected in the field and, consequently, shape it, is our central warning. In fact, *both* difficult-to-detect suboptimality in research products—the lemons—*and* verifiable-high-quality research products—the diamonds—could pass undetected, which would mean that the market for diamonds would become crowded out by lemons. While the crowding out of verifiable-high-quality products in Akerlof's (1970) theorizing leads to adverse selection (whereby all articles would be lemons), this might not directly apply to the entrepreneurship-research market because scholarly articles do not affect each other's 'market success' as in price competition and Akerlof's related mechanisms; yet, we cannot dismiss the possibility of encountering a similar negative outcome.[17] The point is that a whole new generation of AI-supported research articles flooding the entrepreneurship research field deserves vigilant attention because of our current incapacity to detect the lemons *and* diamonds among these articles.

---

[16] Inequality with understanding of AI methods can result from disparities in terms of access to AI. We are in a unique era in the AI race where access to the most powerful AI models are $20 USD a month. This low-cost subscription is unlikely to persist as scholars will all become dependent on them, which will further exacerbate information asymmetry when scholars at well-funded research institutions increase their capacity for research as prices rise and tools become better, while scholars at smaller institutions struggle to keep up.

We acknowledge that the peer-review process is developmental in nature. It is predicated on the idea that scholars reviewing manuscripts are 'experts' in their fields to identify verifiable-high-quality research and help authors improve their work to promote such research. From the used-car market analogy, this is akin to a used-car dealership bringing the used car to a mechanic (even two, three, or four if we count the editor) who can help diagnose the problem, if any, fix it, if possible, and thus figure out whether it is a lemon. The mechanics could thus reduce the information asymmetry and prevent a market for lemons from developing. Our point is that, *at least in the near term*, some mechanics (as a review team) might be unable to diagnose the problem and/or fix it. Consequently, the peer-review process might not ensure that before the 'used car' is offered for sale, the mechanics evaluating its condition have reduced the information asymmetry. When considering both types of information asymmetries (and allied mechanisms) together with the six entrepreneurship scholarship's triggers for a market for lemons, the peer-review process may not deliver on its promises. This issue is further compounded by the six triggers operating not only individually, but also *interdependently* because they can interact with each other. Preventing a market for lemon in our academic field will require a concerted multi-stakeholder effort, and a collective sense of urgency in the field and its knowledge-production and dissemination system (e.g., peer-review process).[18] We next suggest remedies that could prevent such a market—that is, they work against the contributing negative effects. But also, importantly, they work against these detrimental effects to pave the way toward a market for diamonds—verifiable-high-quality research products using advanced AI methods.

[17] Another driver of the increasing lemons crowding out the diamonds could be the specific journal landscape in entrepreneurship scholarship (and business research overall). As highlighted by Grimes et al. (2023), the top journals enjoy a certain exclusive status, and many entrepreneurship scholars must target them for promotion and key performance indicators. Hence, typically only limited publication space for top articles is available, which could limit the chances of diamonds being published if this space is increasingly occupied by (undetected) lemons. On the other hand, we should also note that such increased competition for 'exclusive contracts' (to publish AI-supported research in the field's top journals) could also counter the market-for-lemons dynamics (Rothschild & Stiglitz, 1976). Therefore, the role of top journal exclusivity for potential market-for-lemons dynamics remains unclear.

## 7. Strategic Directions to Avoid a Market for Lemons

One central issue is whether AI methods themselves can facilitate a healthy and productive market for new research findings, and how cutting-edge these methods must be. Can AI itself solve the lemons problem? While some obvious automation potential for AI application exists (e.g., in the peer-review process; see Kankanhalli, 2024), if we follow recent guidelines provided in *Nature* (Bockting et al., 2023:694), then we should *not* delegate all responsibilities to AI, at least in the near term:

> Because the veracity of generative AI-generated output cannot be guaranteed, and sources cannot be reliably traced and credited, we always need human actors to take on the final responsibility for scientific output. This means that we need human verification for at least the following steps in the research process: • Interpretation of data analysis; • Writing of manuscripts; • Evaluating manuscripts (journal editors); • Peer review; • Identifying research gaps; • Formulating research aim; • Developing hypotheses.

We therefore present a list of actionable remedies in Table 1. We acknowledge that this list is rather complex and cumbersome, but this is part of our message: responding to the AI revolution requires deeper cultural changes across various stakeholder groups in our field. Making only one or two adjustments will likely be insufficient; in our view, a more profound update of research infrastructures is needed. This aligns with the notions of stakeholder completeness and collective action in the stakeholder engagement literature (e.g., Bridoux & Stoelhorst, 2022; Lumpkin & Bacq, 2019).

[18] Our multi-stakeholder approach aligns with economic research on the market for lemons highlighting that resolutions should not be limited to a single group active in the market because this group might set standards too low or too high (Leland, 1979).

Our list of remedies was informed by Akerlof's seminal work and the extensive literature on how to best avoid such a suboptimal market. This literature primarily points to the importance of regulation and certification, reputation and trust, transparency and disclosure, market competition, guarantees and warrantees, and third-party verification (e.g., Bochet & Siegenthaler, 2021; Lee et al., 2005; Leland, 1979). Also, ethical issues must be considered when aiming to counteract and/or prevent lemon-saturated markets with these mitigating strategies (Kaplan et al., 2007). For our purpose, mitigating strategies must use standards and checks to ensure that research that uses advanced AI methods is of verifiable high-quality; these strategies could be enabled by institutional regulations to provide a means for countering information asymmetry and, consequently, avoid the creation of lemons. Readers could then be protected and avoid the 'adverse selection' problem whereby the percentage of difficult-to-detect suboptimal articles would increase. In other words, *signaling* must play a central role in helping to fight the market for lemons (Spence, 1973), and so must *screening* (Stiglitz, 2000), whereby (less informed) readers could extract relevant information to evaluate the quality of the research.

For each stakeholder group, we offer broader must-win battles, accompanied by a concrete list of suggested measures to address these battles. However, Table 1 remedies should not be interpreted as exhaustive since future AI developments will most likely prompt adjustments. We also recommend consulting with existing and emerging guidelines and regulations in adjacent fields (e.g., see Gatrell et al., 2024, for management; Landers & Behrend, 2022, for psychology). Moreover, as the US government stepped in with 'Lemon laws' to protect used-car buyers due to the inequality of information between sellers and buyers, we advise consulting AI proposed legislation such as the NYU's Brennan Center[19] in the US that tracks regulations, or the AI Act adopted by the European Parliament to stipulate regulations on "data quality, transparency, human oversight and accountability" (Nahra et al., 2024).

Table 1. Actionable remedies for key stakeholders

| **AUTHORS (academic writers, researchers)** |
|---|
| Must-win battles<br>◊ **Author awareness**: be aware that the double-black-box puzzle (Figure 2) exists<br>◊ **Signal quality**: communicate how the article/book addresses the double-black-box puzzle<br>◊ **Provide transparency**: work against information asymmetries between you (or the AI method) and readers<br>Measures<br>• **Fully implement and follow Open Science Framework** (UNESCO, 2021; see also https://www.cos.io/open-science), when possible: |

[19] https://www.brennancenter.org/our-work/research-reports/artificial-intelligence-legislation-tracker

- Publish data and all codes
- Write research manuscript in a reproducible fashion
- Preregister studies (e.g., registered reports, Chambers & Tzavella, 2022; Open Science: https://www.cos.io/open-science)
- Nurture collectivity/partnership among authors (and their readers): "Collaboration is stronger than competition"

- **Carefully consider and check for bias** in AI results (and in underlying data)—avoid 'blind trust' in algorithms:
  - Consider **AI hallucinations** (i.e., AI output that lacks meaningful or accurate interpretations)
  - Consider all other potential **method biases**, including:
    - **Training data bias**—ask where inherent biases exist in the data used to train the AI model and how representative the data is for what population; the AI model may depend on the quality of the underlying data (garbage in, garbage out)
    - **Feature selection bias**—researchers may unintentionally bias the AI algorithm by selecting features or variables based on their own assumptions or prior beliefs
    - **AI method selection bias**—researcher choses an AI method that is not well-suited to the research question or data characteristics, leading to biased or unreliable results
    - **Evaluation bias**—biased evaluation and interpretation of the AI output
- **Use explainable AI** to overcome the black-box problem of modern AI (Klauschen et al., 2024; Rai, 2020; Taylor & Taylor, 2021)
- **Replication efforts**
  - Consider **direct replication of own AI results** such as a two-study article that includes an original study with AI method and a replication study with non-AI method, or for a research project based on Generative AI using two or more different methods to replicate such findings (Bockting et al., 2023)
  - Consider **using AI for new replication** efforts (Yang et al., 2020)
  - Publish trained machine learning models as **supplementary material** to enable reviewers or future researchers to perform replication studies (and test these models for biases) despite the black box underlying these models
- **Provide additional contextualizing research findings** such as additional qualitative evidence for quantitate AI results
- **Leverage 'cloud intelligence'** by initiating and engaging in **multi-lab studies** (Stephan et al., 2023) and **multi-lab replication efforts** (Dang et al., 2021), but also manage potential quality issues of multi-lab projects (Baumeister et al., 2023).
- **Be mindful of latent field-inherent construct validity issues** such as in study design or results interpretation
- **Use AI to increase research validity** such as scale development (Götz, 2023)
- **Do not throw established theory overboard** and engage in 'bold' theory building and testing (Lévesque et al., 2022)
- **Invest in personal AI savviness** but still communicate studies in a way that readers with less AI knowledge can still follow (by presenting examples and defining terms and concepts) and carefully check the quality of the research
- **Be aware of your own and collective biases** and other cognitive limitations (Tversky & Kahneman, 1974, 1991) by considering that "proposed AI solutions can also exploit our cognitive limitations, making us vulnerable to illusions of understanding in which we believe we understand more about the world than we actually do" (Messeri & Crockett, 2024:49)
- **Avoid the Clever Hans effect**[§] where the researcher, or the researcher and the audience, unintentionally produce a desired research outcome (Lapuschkin et al., 2019)
- **Adopt human safeguarding** by preparing for when advanced AI might outperform researchers in entrepreneurship research capacities but assume that guarding the quality and reputation of the research field may remain primarily the responsibility of (human) entrepreneurship scholars (Landers & Behrend, 2022) or government or industry regulatory agencies
- **Engage in resource acquisition** such as fundraising/grants, lobbying with stakeholders, interdisciplinary collaboration, to access, maintain, and audit research resources required for high-quality, transparent AI-supported research, such as AI infrastructure and knowledge

**REVIEWERS**

Must-win battles

◊ **Reviewer awareness**: be aware that the double-black-box puzzle (Figure 2) can mislead
◊ **Request transparency**: work against information asymmetries between the author(s) and reviewers
◊ **AI resilience**: invest in AI training—reviewers play a central role in detecting lemons at an early stage

Measures

- **Same remedies as for authors** (e.g., request authors to apply the abovementioned remedies)
- **Be sensitive to, and work against, information asymmetries** by requesting thorough information from authors regarding the data and AI analysis and the evaluation and interpretation of the AI results, and aim to achieve a level playing field, putting you, the reviewer, in the authors' shoes
- **Encourage third-party verification** by encouraging authors to undergo third-party validation of their research findings (by reputable organizations or experts who can enhance research credibility and reduce the risk of lemons)
- **Use AI with great care and transparency** in the review process (and only if allowed by the journal[§§])—be mindful that the black-box problem of modern AI could also lead to new information asymmetries in the review process when using AI for peer-reviews
- **Help the editor** by signaling if you, as an author, can effectively evaluate the adequacy of the AI method employed, or whether you wish to leave this to the editor, or recommend that the editor consult an AI method expert
- **Be aware of the cognitive bias overconfidence** that can fuel the market for lemons (e.g., by overestimating one's ability to estimate the quality of a research article, Herweg & Müller, 2016)

**EDITORS**

Must-win battles

◊ **Regulatory oversight**: provide regulations/standards to ensure quality of published articles using AI methods
◊ **Trust**: increase and maintain trust between authors and readers
◊ **Educate**: help educate reviewers on how to assess articles using AI methods

Measures

- **Same remedies as for authors** (e.g., request authors to apply the abovementioned remedies, value replication)
- **Consider Open Access** and **Open Science initiatives**
- **Publish clear, detailed guidelines and regulations** on how authors, reviewers, and editors should use AI (Brainard, 2023; Flanagin et al., 2023; Gatrell et al., 2024)
- **Form a taskforce** focused on creating guidelines and regulations for the journal that are sensitive to the field's unique evolution and *status quo*
- **Work proactively against information asymmetries** (e.g., between author and reviewer/editor) by considering how reviewers and authors communicate with each other in the peer review and revision process
- **Consider recruiting additional associate editors** familiar with using advanced AI methods
- **Create certification systems** by introducing peer endorsements for research articles that meet certain quality criteria (this can assure readers that the research has undergone rigorous scrutiny and meets established standards)
- **Introduce visible 'quality badges'** on papers following the Open Science Framework (Schneider et al., 2022) (e.g., for preregistration or open data, see also https://www.cos.io/initiatives/badges)
- **Value replication studies** as they define how proper scientific research is conducted (e.g., in medicine)
- **Work against reviewers' dissatisfaction** (e.g., if they feel less competent to review AI studies, but are increasingly asked to)
- **Promote special issues** in the AI space (e.g., Obschonka & Audretsch, 2020; Obschonka et al., 2024), which could also emphasize ways to prevent information asymmetry (e.g., implement guidelines suggested for authors, reviewers, and editors in this table)
- **Audit the AI auditors** by using regular checks on: 1) the submission and review guidelines; and 2) how authors and reviewers use them (Landers & Behrend, 2022)

**PUBLISHERS / JOURNALS**

Must-win battles

◊ **Support and advise**: all stakeholders involved to identify and root out a market for lemons
◊ **Innovation**: consider innovative measures and solutions to help all stakeholders to identify lemons

Measures

- **Choose quality over quantity** in research publications
- **Quality standards and guidelines** for establishing and enforcing quality standards and guidelines for submitted manuscripts (e.g., criteria for methodological rigor, ethical conduct, and reporting transparency)
- **Support editors** in prioritizing the development of journal-/field-specific AI regulations and standards for authors, editors, and reviewers
- **Provide advice** on the pros and cons of automating the review process using AI (Checco et al., 2021)

- **Identify wider information asymmetries** in the review process and in the dissemination of published articles such as asymmetries between authors, reviewer, and editors on the one side, and readers of the published work on the other side—promote transparency in the peer review process
- **Implement quality certification** by considering programs where articles using advanced AI methods are certified by independent organizations or regulatory bodies, because certification helps signal quality to readers and reduces the risk of 'consuming' low-quality articles
- **Introduce warranties and guarantees** by implementing quality assurances for articles using AI methods to reassure readers about the quality of such articles
- **Avoid crowding out of diamonds** by, for instance, making sure that new generations of published articles using AI methods that are of unclear quality do not crowd out high-quality articles because of limited journal space (e.g., by promoting special issues or other types of extra publication space)
- **Train Generative AI algorithms** for scholarly searches so that these algorithms can flag the market-for-lemon dynamics that can produce knowledge loops that can impact new research, and reinforce the broader harms caused by the AI revolution
- **Provide training and recognition for reviewers** to ensure they have the skills and knowledge to effectively assess the quality of submitted manuscripts
- **Promote post-publication peer review** by offering mechanisms for post-publication peer review, such as commenting systems or letters to the editor, to allow for ongoing feedback, evaluation, and discussion of published research

**EMPLOYERS / UNIVERSITIES**

Must-win battles

◊ **Capacity building**: provide AI resources and training
◊ **Incentivize**: motivate scholars to implement recommendations suggested herein
◊ **Ethics**: provide legal and ethical guidance

Measures

- **Empower entrepreneurship researchers** to produce and disseminate AI-supported research at the highest level by providing AI infrastructure and other required resources, including the training and promotion of collaboration with AI experts such as data scientists (Jones, 2023b)
- **Reward open-science practices** in entrepreneurship scholars (e.g., via KPIs or research incentive programs)
- **Invest in AI education of PhD and ECRs**
- Help to **democratize AI** by giving university scholars access to advanced AI infrastructure (comparable to researchers in major commercial labs), facilitating research partnerships with commercial labs (Lundvall & Rikap, 2022)
- **Promote collaboration and mentorship** by encouraging collaboration, particularly interdisciplinary collaboration and mentorship among faculty members, researchers, and students
- **Reward quality over quantity** by evaluating scholars based on the quality and impact of their research outputs rather than solely on quantitative metrics such as publication counts or grant funding, and recognize and reward scholars who produce rigorous, innovative, and impactful research
- **Support research integrity and research ethics boards** by providing AI-resilient and effective research ethics boards that advise authors on ethical standards for research that use advanced AI methods (e.g., by reviewing research proposals)

§ The Clever Hans effect originated from a horse, Clever Hans; in the early 20th century, Clever Hans appeared to perform arithmetic calculations but was responding to subtle cues from his trainer and audience (Pfungst, 1911). This effect is often cited when discussing potential biases in modern AI research and application (e.g., Klauschen et al., 2024).
§§ For instance, Gatrell et al. (2024:9) argue that generative AI "has no place in the peer review process" (see also Brainard, 2023).

Before concluding, we must highlight that the new AI era, translated into fascinating practice-oriented questions (e.g., how entrepreneurs use AI to scale their business startups), should not distract from the important need to maintain high research standards and facilitate effective knowledge exchanges. Additionally, we must keep an eye on both methodological fundamentals and 'healthy' methodological advancements (e.g., using AI methods in innovative ways), which will require taking some risks in how we use and develop new research methods and processes. However, some argue that entrepreneurship scholars might, somewhat paradoxically, often favor relatively conservative and risk-averse research approaches that are less 'bold and entrepreneurial' (see Lévesque et al., 2022). As highlighted by Shepherd (2015:489), this could have to do with a competency trap

> …that rewards in the short run playing it safe by using 'accepted' theories and approaches to address increasingly narrow research questions of interest to a smaller audience. This is not to say that this type of incremental research is not important to the field. However, if incremental research begins to dominate and crowd out more transformational research, we run the very real risk that the field will stagnate and lose what is 'special' about it.

While these lines are nearly a decade old, they might still hold true today. Although we have witnessed many developments (e.g., many highly useful method-focused editorials and more formal and ambitious methods standards), the field might still need to develop a new balance between research innovation and protecting its unique core and orientation—a phenomenon-oriented, practice-based interdisciplinary field, particularly in the unfolding AI era.

## 8. Conclusion

Entrepreneurship is a fascinating, timely, and critical research topic, addressed by a broad research community. This research field has a deeply rooted upward trajectory, though at times with meandering paradigm discussions that are intellectually exciting but scientifically complex. The field has become a vibrant but knotty marketplace in terms of size, global representation, and scholarly exchanges (Landström & Harirchi, 2018). We attempted to make a call to action, individually and collectively, to ensure optimal knowledge exchanges. Although some might not share our optimism, we believe that the field can renew the enthusiasm and sense of departure that accompanied its golden era at the turn of the 21st century. We hope we can make a small contribution to propelling the field ahead of the curve, instead of merely reacting to technological progress. This proactive approach should serve us well because, as entrepreneurship scholars, we should feel more at home with proactive innovation and acting—much like *Facebook*'s former entrepreneurial motto: "Move fast and break things"—compared to passively waiting and hoping that the AI revolution *will not* break things in our field like productive knowledge exchanges, rigor, and reputation (*Nature Machine Intelligence*, 2023). By achieving a level of AI resilience that does not leave us continuously fearing whether it breaks things or not (e.g., optimal knowledge production and exchanges), we can secure the confidence, boldness, and trust to master and leverage this unprecedented revolution. As highlighted by Lévesque et al. (2022), let's give *us* agency and control, let's make AI *our* servant, collectively as a field and with vigilance.